\documentclass[useAMS,usenatbib]{mn2e}
\usepackage{amsmath,amssymb}
\usepackage{graphicx,mathptm}
\usepackage{times}
\usepackage{natbib}

\title[Hierarchical evolution of AGNs]
{Modeling active galactic nuclei: ongoing problems for the faint-end of the luminosity function}

\author[F.~Marulli et al.]
  {F.~Marulli$^1$, E.~Branchini$^2$, L.~Moscardini$^{1,3}$ and  M.~Volonteri$^4$ \\
  $^1$Dipartimento di Astronomia, Universit\`a degli Studi di Bologna,
      via Ranzani 1, I-40127 Bologna, Italy \\
  $^2$Dipartimento di Fisica, Universit\`a degli Studi ``Roma Tre'',
      via della Vasca Navale 84, I-00146 Roma, Italy\\
  $^3$INFN, Sezione di Bologna, viale Berti Pichat 6/2, I-40127 Bologna, Italy\\
  $^4$Institute of Astronomy,  Madingley Road, Cambridge, CB3 0HA, U.K. \\
 }
\pagerange{\pageref{firstpage}--\pageref{lastpage}}
\pubyear{2005}

\begin{document}

\label{firstpage}
 
\maketitle

\begin{abstract}

We consider simple semi-analytic models 
that relate the active galactic nuclei (AGN) evolution to the merging history 
of their host dark matter haloes and quantify their ability of matching 
the AGN luminosity function and its spatial clustering at low and intermediate redshifts. 
In particular, we focus on the recent determinations of the AGN 
luminosity function in the hard X-ray band at $z\sim0$ which
constitutes the most stringent observational test for our models.
Indeed, while we find an acceptable agreement between the model bolometric luminosity 
function and the data at $1\lesssim z\leq2$ and for luminosities larger than $10^{10}L_{{\rm bol},\odot}$, 
no semi-analytic model is capable of reproducing the number density of faint 
X-ray sources in the local universe.
Some improvement can be obtained by advocating energy feedback that we
model through a time-dependent Eddington ratio. Even in this case, however, 
the number density of faint AGNs is significantly below observations.
This failure indicates that major mergers cannot constitute the only trigger 
to accretion episodes in the local AGN population.

\end{abstract}

\section{Introduction}

The luminosity function (LF) of AGNs, 
namely the derivative of their co-moving number 
density with respect to luminosity,
is a very important tool
to understand their cosmological evolution. In the recent
years, the AGN LF has been measured in a wide range of redshifts, luminosities
and in different wavelength bands: radio \citep[see e.g.][]{nagar2005},
optical \citep[see e.g.][] {kohler1997,grazian2000,croom2005,richards2005,richards2006,siana2006}, 
infra-red \citep[see e.g.][] {brown2005,matute2006}, 
soft X-ray \citep[see e.g.][]{miyaji2001,hasinger2005}, hard X-ray \citep
[see e.g.][]{ueda2003,sazonov2004,lafranca2005,shinozaki2006,beckmann2006}.
In this work we are mainly interested in the two most recent 
determinations of the AGN LF 
in the hard ($\ge 2 {\rm keV}$) X-ray band that, despite being very local,
provides strong constraints to AGN models.
The first one, provided by \citet{shinozaki2006} (hereafter S06), 
consists of a complete, flux-limited sample of 49 
sources  from the HEAO-1 All-Sky catalogue, complemented with spectral information 
from ASCA, XMM-Newton and Beppo-SAX observations.
All objects in the catalogue
are optically classified as emission-line Seyfert galaxies
at high galactic latitude ($b\ge 20^{\circ}$) 
with column density $N_H>10^{21.5}{\rm cm^{-2}}$ and $L_X=L[2-10 {\rm keV}]>10^{42}\,\rm erg \, s^{-1}$. 
The second AGN LF has been determined in a harder X-ray band $[20-40 {\rm keV}]$ 
by \citet{beckmann2006} (hereafter B06), using a sample 
of 38 objects, preferentially located at low galactic latitude,
detected by the imager IBIS/ISGRI on-board INTEGRAL,
with $L_X=L[20-40 {\rm keV}]>10^{41}\,\rm erg \, s^{-1}$.
The main reason for concentrating on these two datasets is that
they allow to span a large band $[2-40 {\rm keV}]$ in which it is possible to
detect absorbed and unabsorbed AGNs at all galactic latitudes.

Further observational constraints to theoretical models 
are provided by the AGN spatial clustering, 
which is often quantified by means of the angular or spatial two-point correlation function.
Uncertainties in current modeling of the AGN clustering, however, 
make this second constraint less effective than the LFs. 
In spite of that, in this work we also 
check the ability of our models in matching the AGN biasing function,
defined as the ratio between the spatial two-point correlation function 
of AGNs and dark matter (DM): $b^2(M,z)=\xi_{AGN}/\xi_{DM}$, 
where $M$ is the mass of the hosting halo.
Also in this case, we consider the very recent determination of the
AGN  biasing function in the B-band by \citet{porciani_norberg2006} (hereafter PN06),
which provide the most accurate estimate of the AGN clustering to date.

Over the years, several analytic \citep[see e.g.][]{efstathiou1988,haehnelt1993,haiman1998, 
percival1999, haiman2000,martini2001, hatziminaoglou2001,wyithe2002,wyithe2003,hatziminaoglou2003} 
and semi-analytic \citep[see e.g.][]{Cattaneo1999,kauffmann2000,cavaliere2000, 
cattaneo2001,cavaliere2002,enoki2003,volonteri2003a,springel2005, 
cattaneo2005,croton2006,malbon2006} models have been proposed to
describe the co-evolution of supermassive 
black holes (BH) powering AGNs and their DM halo hosts 
within the hierarchical clustering framework.
These models  proved to be successful in matching the
optical LF of AGNs around their peak of activity at $z \sim 2$.
However, modifications to the simple analytic models in which the AGN activity is triggered by 
haloes' mergers have been subsequently introduced 
to bring predictions into agreement with the observed AGN LFs both at higher and lower redshifts.
For example, selective BH accretion at early stages occurring either at super-critical rate
\citep[e.g][]{volonteri_rees2005}
or at Eddington rate for BHs hosted in the high-density peaks \citep[e.g.][]{volonteri_rees2006},
can explain the observed number density of AGNs at $z \simeq 6$.

At low redshifts ($0.5<z\le2$), inefficient cooling in large haloes 
is required to improve the match to the bright end of the LF in both 
the optical and hard X-ray bands \citep{marulli2006} (hereafter M06).
In the same redshift range, \citet{volonteri_salv_haardt2006}
have recently shown that inefficient accretion, suggested by the
results of numerical simulations that track accretion onto BHs 
following halo mergers \citep{hopkins2005},   
increases the number density of faint AGNs matching their 
optical luminosity function.
Whether these modifications can
also reproduce the LF of AGNs in the very nearby universe
and in other luminosity bands it is still matter of debate.

Before addressing this question, it is worth stressing that
comparisons between model and observed LFs are hampered by the fact that, while the former
refer to bolometric luminosities, the latter are measured in some specific bands. 
Luminosity band corrections 
represent therefore a key issue that has been addressed by 
several authors \citep[see e.g.][]{elvis1994,marconi2004}. 
In this paper, we adopt the bolometric correction recently proposed by \citet*{hopkins2006} (hereafter H06)
using a number of  determinations of AGN LFs in the interval $0<z\leq5$ 
and in different bands ranging from 
infrared to optical, soft and hard X-ray bands. 
It is worth noticing that the H06 bolometric correction was not calibrated using the
S06 and B06 results. Yet, as we have verified, the 
bolometric correction successfully applies to both LFs. 
This is less obvious for the B06 than for the S06 dataset
which, in fact, contains a significant fraction of objects that
have been included in the H06 analysis.

The main aim of this work is to predict the LF of AGNs using
standard semi-analytic hierarchical models at $z\sim 0$
and compare it to the S06 and B06 results by using the 
H06 bolometric corrections. In doing so, we also
show model predictions for the AGN LFs and clustering 
up to $z=2$, hence updating the results of M06.

The outline of the paper is as follows. 
In the first section we briefly present the semi-analytic models
considered in this work and summarize 
the main assumptions used therein.
In Section 3 we compare model predictions with the observed luminosity 
and biasing function of AGNs.
Finally, in the last section we discuss our results and draw our main
conclusions.

Throughout this paper we assume a flat $\Lambda$CDM cosmological model
with Hubble constant $h\equiv H_0/100\,{\rm km\,s^{-1}}\,{\rm Mpc}^{-1}=0.7$, a
dominant contribution to the density parameter from the cosmological
constant, $\Omega_{\Lambda}=0.7$, and a CDM density power
spectrum with primordial spectral index $n=1$.

\section{Semi-analytic models}
\label{sec:models}

Most of the semi-analytic models used in this 
work have been already considered by M06.
They all assume that the evolution of AGNs can be followed
within the standard hierarchical scenario of structure evolution 
under the hypothesis that, at every redshift, 
the AGN activity is solely determined by the cosmological merging history of 
their DM halo hosts. In this framework, the merging history 
of dark haloes is described by the extended Press \& Schechter 
formalism \citep{bond1991,lacey1993}, while phenomenological prescriptions 
are adopted to model the feeding of the central BH 
and  the physical processes of the AGN activity.

Like in M06, the models are based on the semi-analytic code developed by \citet*{volonteri2003a}.  
AGN activity is assumed to be triggered  when {\em major mergers} occur, i.e. when 
the two merging haloes have a mass ratio larger than $0.1$. 
Due to the lack of an exhaustive study of the ultimate consequences of a
galaxy merger in its whole parameter space, we are forced to make 
some simplifying assumptions to follow the merging events.
Following \citet{cox2004}, we can assume that all halo mergers,
except the ones with mass ratio smaller than $0.1$, can destabilize the 
gas at the centre of the more massive halo, and consequently
induce star formation and BH mass accretion.
Notice that this threshold is lower than the value of $0.3$ generally used 
in the literature, but it is not low enough to reproduce the observed 
faint AGN number counts, as we will describe later.
So, an higher value of the mass ratio would worsen our results,
while a lower one would be in disagreement with the results of \citet{cox2004}, 
which show that a typical merger with a mass ratio of $~0.05$ does
not induce starbursts.
Moreover, according to \citet{taffoni2003}, when $P<0.1$ the dynamical
friction timescale is larger than the Hubble time,
hence preventing the merging of satellite galaxies and, reasonably,
making the accretion efficiency onto the central BH very low.

We stress here that considering galaxy rather than halo mergers could 
lead to different predictions as it is not always true that a major merger
of two dark matter haloes will result in a major merger of their galaxies
(and viceversa). A galaxy-merger-driven scenario has the virtue of matching
the observed correlation between BHs and their galaxy hosts \citep{magorrian1998,
ferrarese2000,gebhardt2000} but can only be self-consistently
implemented within the framework of a full semi-analytic or numerical model of
galaxy formation and evolution. This is beyond the scope of this work in which, 
instead, we use a  model meant to minimise the number of free parameters.

It is also assumed that the seed BHs formed with masses of $150\,M_\odot$ 
following the collapse of the very rare Pop III stars, in minihaloes
forming at $z=20$ from the density peaks above a $3.5\sigma$
threshold; however, as
shown by \citet{volonteri2003a}, the final results are not very sensitive to this choice. 
After every major merger, the BH at the centre of the more massive
halo increases its mass, $M_{\rm BH}$, after a dynamical
free-fall time when a significant fraction of the gas falls to the
centre of the merged system (\citealt*{springel2005b, dimatteo2005}) and is
accreted at an appropriate rate. To implement this mechanism
we need to specify the prescriptions for the mass accretion,
that can be modeled by the following relation
\begin{equation}
  M_{\rm BH}(t+\delta t)=M_{\rm BH}(t)\exp\left(\int\frac{\delta t}{t_{\rm Edd}}
  f_{\rm Edd}(t)\frac{1-\epsilon}{\epsilon}\right),
  \label{eq:1}
\end{equation}
in which $t_{\rm Edd}=0.45\,{\rm Gyr}$, $f_{\rm Edd}$ is the Eddington ratio of mass accretion,  
and $\epsilon$ is the radiative efficiency. 

We have considered four different scenarios.
\begin{itemize}
\item
In the first one, labeled E1 in all plots, the accreted mass is
proportional to the mass of the available gas and hence to the total
mass of the more massive progenitor: $\Delta M_{\rm accr}=\alpha M_{\rm
halo}$. We set $\alpha=2\times 10^{-5}$, in agreement with
the normalization of the $M_{\rm BH}-\sigma_g$ relation at $z=0$, where $\sigma_g$ is the
velocity dispersion of the host galaxy \citep{tremaine2002}, scaling
with the halo circular velocity, $v_c$, as suggested by
\citet{ferrarese2002}.
As $M_{\rm halo} \propto v_c^3$, the slope of
the $M_{\rm BH}-\sigma_g$ relation is flatter than the observed one
\citep[but see][]{wyithe2005a}. 
\item
The second scheme, labeled E2, 
assumes a scaling relation between the accreted
mass and the circular velocity of the host halo, $\Delta
M_{\rm accr}\propto k\cdot v_c^5$, which is normalized {\it a-posteriori} to
reproduce the observed relation between $M_{\rm BH}$ and $v_c$ at $z=0$ 
\citep{ferrarese2002}. As in M06, we assume a linear dependence of $k$
on redshift, as $k(z)=0.15(1+z)+0.05$, in order to account for the decrease
of the gas available to fuel BHs. Unlike model E1, here and in the other three models
the relation between $M_{\rm BH}$ and $M_{\rm halo}$ evolves in redshift
as in \citet{wyithe2003}: 
$M_{\rm BH}\propto M_{\rm halo}^{5/3}\cdot(1+z)^{5/2}\cdot(\Delta_c/\Omega_m(z))^{5/6}$,
in which $\Delta_{c}(z)=18\pi^{2}+82d-39d^{2}$, $d\equiv \Omega_{\rm
m}(z)-1$ and $\Omega_{\rm m}(z)$ represents the mass density
parameter.
Finally, we  account for inefficient cooling 
in large haloes by preventing accretion 
within haloes of masses $M_{\rm halo}>10^{13.5} M_{\odot}$.
It is worth noticing that this prescprition
has a physical motivation connected to both galaxy and AGNs formation since it
has the same effect of including the low luminosity radio mode
AGN heating, as done in many semi-analytic models of galaxy formation to produce a massive galaxy
population similar to the one observed \citep[see e.g.][]{kang2006,bower2006,Cattaneo2006,croton2006}.

\item
Another model, labeled B, assumes an early
stage of super-critical accretion during which the central BH (AGN) accretes
mass at a rate that can be estimated by the Bondi-Hoyle formula \citep{bondi1944}. 
This model applies to metal-free haloes, therefore
we assume that by $z=12$ the interstellar medium has been enriched,
and we inhibit super-critical accretion rates. When the super-critical
phase ends, accretion proceeds in subsequent episodes as in model E2.
This possibility has been recently advocated by
\citet{volonteri_rees2005} 
to reconcile a hierarchical evolution with
the existence of QSO at $z\sim 6$, hosting SBHs with masses $\sim 10^9
M_\odot$.

In all these models, which do not account for possible feedback mechanisms, 
we set in Eq. (\ref{eq:1}) the Eddington ratio of mass accretion, $f_{\rm Edd}\equiv \dot{M}/\dot{M}_{\rm Edd}=1$. 
\item
In this work we consider also a different model, labeled H, which 
accounts for the results of the recent hydrodynamic simulations
of galactic mergers in which AGN feedback is included \citep{hopkins2005}
that show that the Eddington ratio is not constant but depends on AGN luminosity.
As the main variable in the model is the BH mass rather than 
the AGN luminosity, \citet{volonteri_salv_haardt2006} model $f_{\rm Edd}(t)$ as follows.
First the time spent by a given AGN per logarithmic luminosity interval is approximated as
\begin{equation}
  \frac{dt_{\rm AGN}}{d\log L}=|\alpha|t_Q\left(\frac{L}{10^9L_\odot}\right)^\alpha,
\end{equation}
where $t_Q\simeq10^9\,yr$, $\alpha=-0.95+0.32\log(L_{\rm peak}/10^{12}L_\odot)$, 
and $L_{\rm peak}$ is the luminosity
of the AGN at the peak of its activity and it can be approximated as the Eddington luminosity
of the BH at its final mass, i.e. when it sets on the $m_{\rm BH}-\sigma_g$ relation \citep{hopkins2005}.
Then, since the AGN luminosity can be written as 
$L=\epsilon f_{Edd}(t)\dot{M}_{\rm Edd}c^2$, where $\epsilon$ is the radiative efficiency, 
$\epsilon=L/(f_{\rm Edd}(t)\dot{M}_{\rm Edd}c^2)=0.1$
\footnote{The radiative efficiency has been self-consistently determined by tracking the evolution 
of BH spins throughout the calculations \citep{volonteri2005}.},  
the following differential equation is used to describe the evolution of  $f_{\rm Edd}(t)$ 
\begin{equation}
  \frac{df_{\rm Edd}(t)}{dt}=\frac{f_{\rm Edd}^{1-\alpha}(t)}{|\alpha|t_Q}
  \left(\frac{\epsilon \dot{M}_{\rm Edd}c^2}{10^9L_\odot}\right)^{-\alpha}.
  \label{eq:fedd}
\end{equation}
The instantaneous Eddington ratio $f_{\rm Edd}(t)$ is obtained by solving Eq. \ref{eq:fedd}.
Model H assumes that the final mass of the black hole is determined by 
the circular velocity of the host halo, as in model E2. 
\end{itemize}

\section{Models vs. observations}
\label{sec:MovsObs}

\subsection{The bolometric luminosity function between $0.5\le z\le 2$}

\begin{table*}
  \begin{center}
    \caption{Values of $\Xi^{\rm model}$ and the corresponding per cent probability $P(\Xi^{\rm zamodel})$, indicated in parentheses,  
     for each models and redshifts.} 
    \label{tab1}
    \begin{tabular}{cccccc}
      \hline
      \hline
      Model &                  &                  & $\Xi\,(P(\Xi))$   &                  &       \\
            & $z=0.1$          & $z=0.5$          & $z=1$         & $z=1.5$          & $z=2$ \\
      \hline
      E1 & 5.3 (0.2\%) & 2.2 (13.6\%) & 1.5 (34.9\%) & 1.1 (45.0\%) & 6.6 (2.5\%)  \\
      E2 & 2.5 (2.4\%) & 3.3 (7.4\%)  & 4.6 (4.9\%)  & 2.8 (16.4\%) & 0.5 (80.8 \%)\\
      B  & 5.4 (0.3\%) & 5.2 (3.2\%)  & 5.7 (4.0\%)  & 3.1 (12.2\%) & 1.1 (50.8 \%)\\
      H  & 2.3 (2.5\%) & 3.7 (6.0\%)  & 5.5 (4.3\%)  & 1.9 (26.9\%) & 0.4 (90.8 \%)\\
      \hline
      \hline
    \end{tabular}

  \end{center}
\end{table*}

\begin{figure*}
\includegraphics[width=0.48\textwidth]{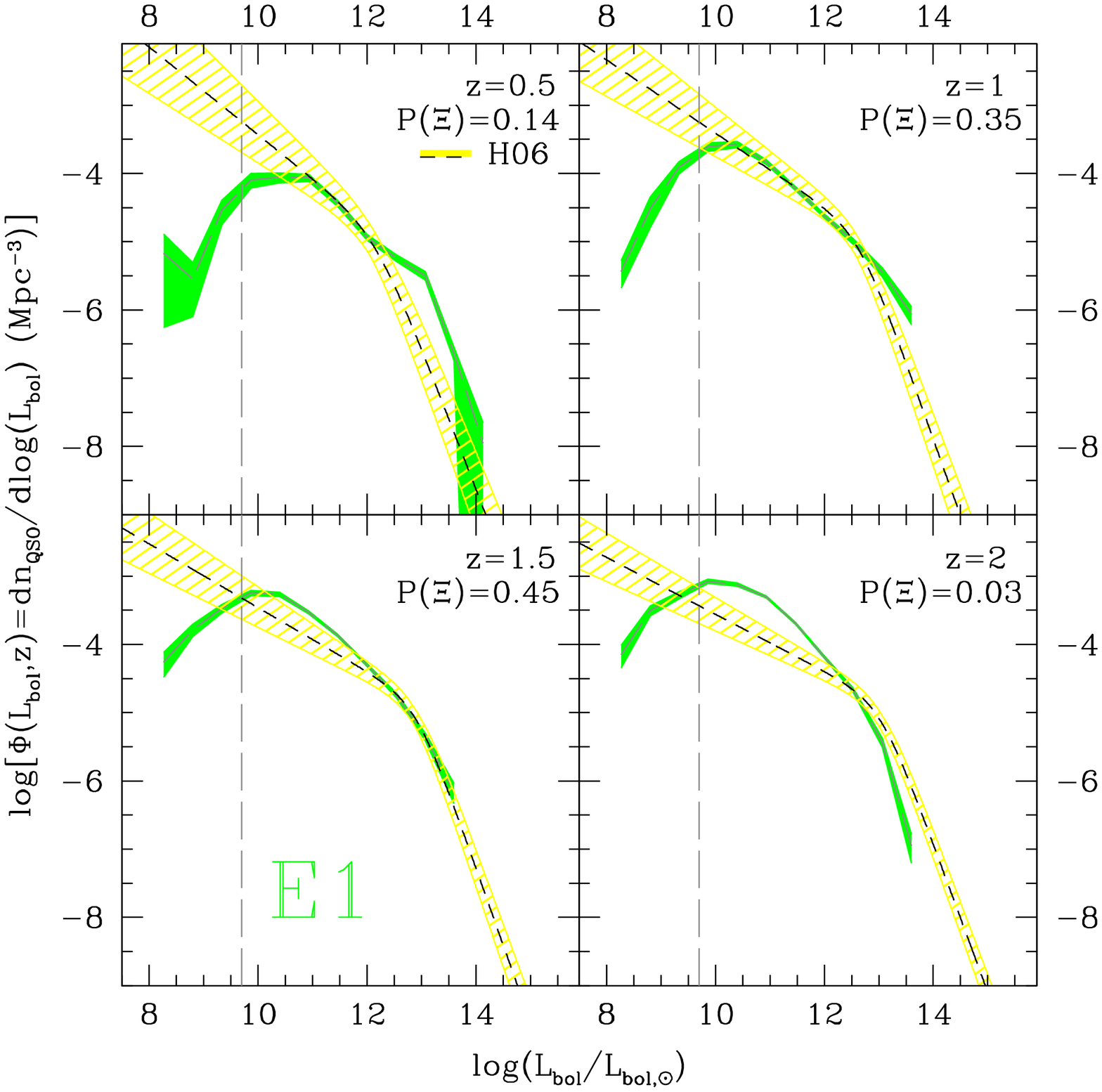}
\includegraphics[width=0.48\textwidth]{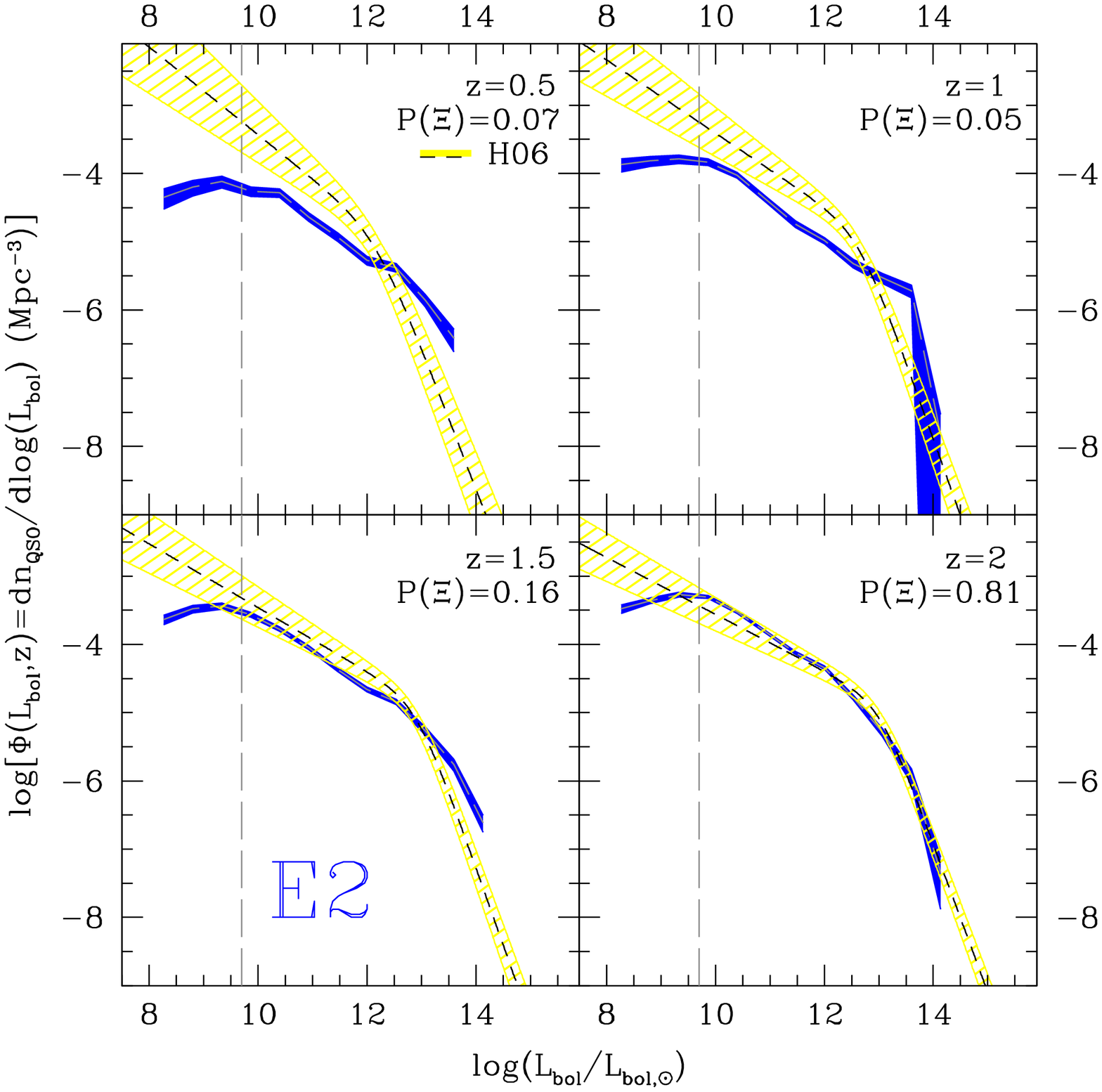}
\includegraphics[width=0.48\textwidth]{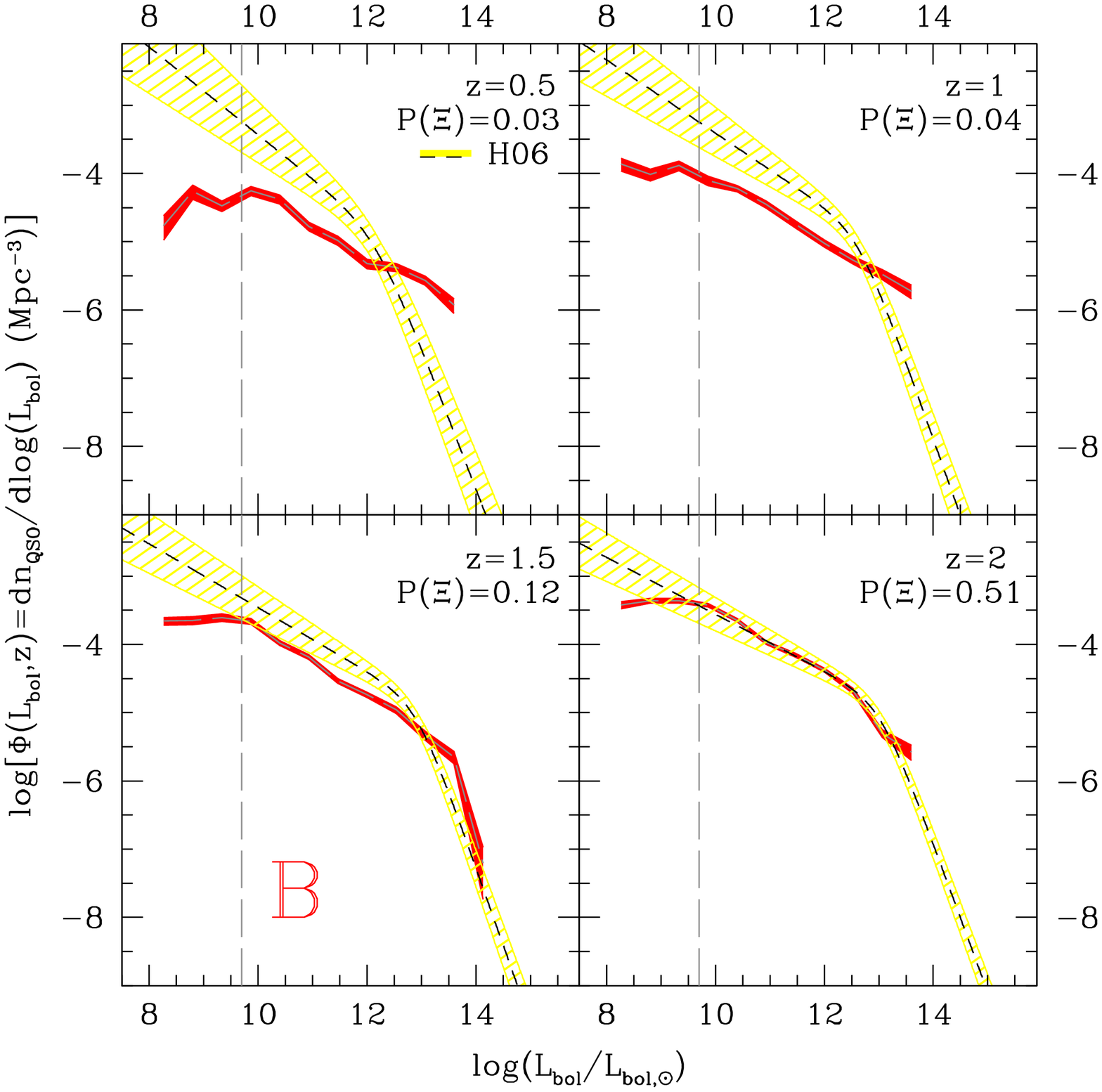}
\includegraphics[width=0.48\textwidth]{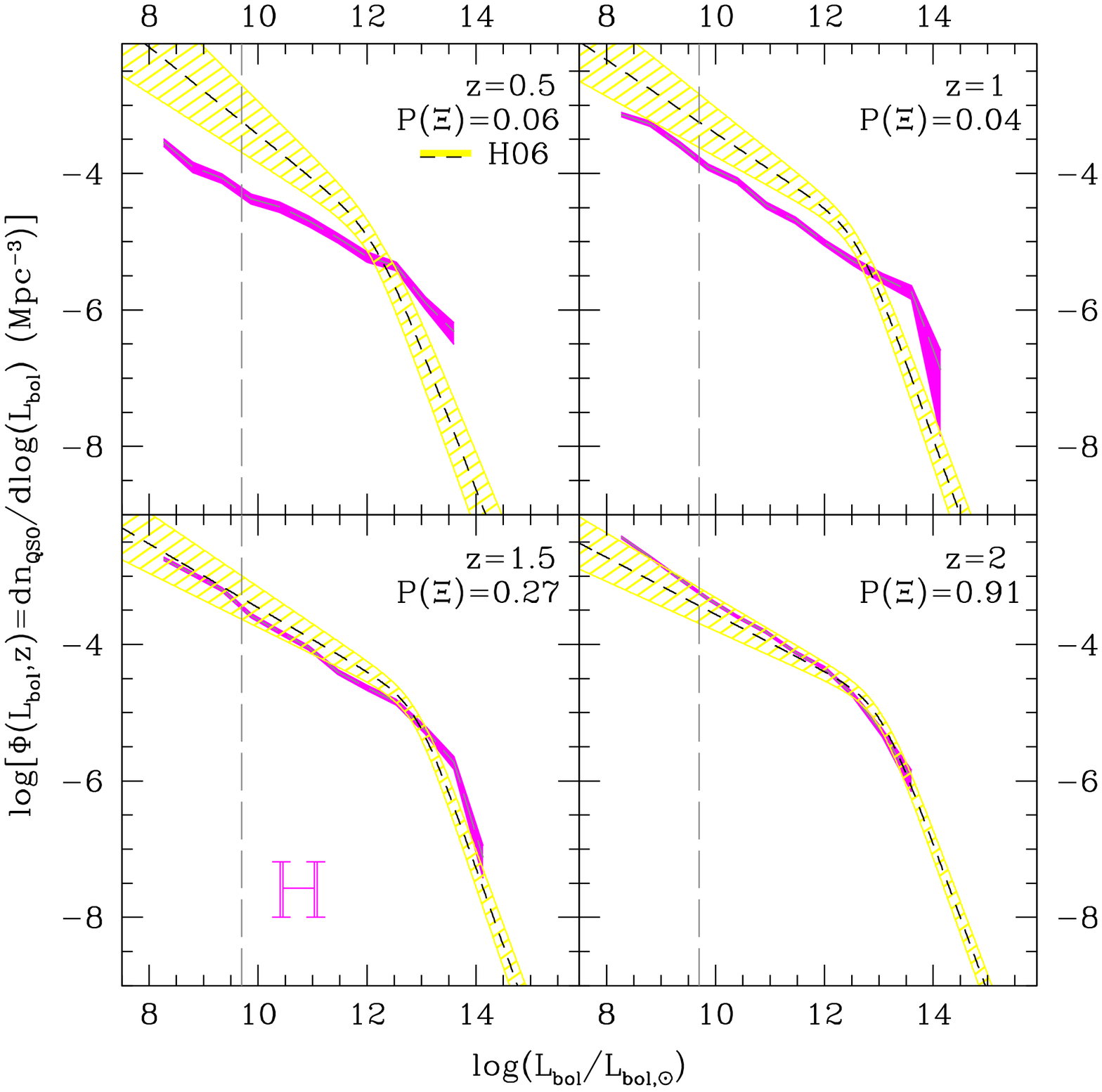}
\caption{
The AGN bolometric LF at $0.5\leq z\leq2$: models vs. observations. 
The dashed black lines show the bolometric LF of H06, while the yellow shaded areas 
take account of the estimated errors of the fit. 
The dashed vertical lines show the minimum bolometric luminosity
accessible to observations. 
Each set of plots, composed by four panels corresponding to different redshifts, 
refers to a different model, as indicated by the labels.
}
\label{fig:lf1}
\end{figure*}

To compute the model LF we have considered binary merger trees
with masses in the range $(1.43\times10^{11}M_\odot, 10^{15}M_\odot)$  
\citep{volonteri2003a}.  M06  have used  models E1, E2 and B described in the 
previous section to follow the accretion history of the BHs 
and the evolution of AGN luminosity at the centre of haloes.
Here we have repeated this procedure to implement the novel model H .

We have calculated the bolometric LF by
simply discretizing the luminosity range of our modeled AGN sample 
in finite bins, counting their number 
density in all our merger trees, each of them weighted by the
halo number density at $z=0$ \citep{sheth1999}
and normalizing for the number of merger trees and 
time-steps considered. Uncertainties in the model LF have been computed
assuming Poisson statistics.

As a first step, we extend the M06 analysis by comparing the bolometric luminosity
function of H06 with the analogous quantity predicted by the four models
at redshifts $z=0.5, \ 1.0, \ 1.5$ and $2$. 
This comparison represents a more severe test to the models  
than the one performed by M06, since the number of AGNs used by H06
to model their bolometric LF is significantly larger than those
considered by M06, and consequently the uncertainties are smaller.
The results are shown in Fig. \ref{fig:lf1}, which is divided in four sets 
of plots, each one referring to a different model.
In each set, composed by four panels, the dark-shaded area represents the $1\sigma$ 
uncertainty strip around the model LF. 
The dashed curve shows the bolometric LF of H06 along with $1\sigma$ uncertainty strip, plotted 
as a light-shaded area. In all plots the vertical, dashed line shows the minimum bolometric luminosity 
accessible to observations, $L_{\rm obs}^{\rm min}$,
which turns out to be remarkably constant in the interval of redshifts considered.
Model predictions extend up to a maximum luminosity 
$L_{\rm model}^{\rm max}\sim10^{14} L_\odot$
resulting from having set an upper limit to the mass of our DM haloes
($M^{\rm max}\sim10^{15}M_\odot$).

To quantify the consistency between models and data we have estimated
the following $\chi^2$-like quantity:
\begin{equation}
 \Xi^{\rm model}(z)=\frac{1}{N_{\rm bin}}\sum_{i=1}^{N_{\rm bin}}\frac{\left
[\log(n_{\rm model}({\Delta L_i},z))
     -\log(n_{\rm obs}({\Delta L_i},z))\right]^2}{\sigma_{\rm model}^2+
\sigma_{\rm obs}^2}\ ,
\end{equation}
where $n_{\rm model}({\Delta L_i},z)$ and $n_{\rm obs}({\Delta L_i},z)$
represent the model and observed mean comoving number density of AGNs
in the luminosity interval $\Delta L_i$ at redshift $z$, $\sigma_{\rm model}$ and
$\sigma_{\rm obs}$ are the  $1\sigma$ logarithmic errors 
and the sum runs over the
$N_{\rm bin}$ luminosity bins in the interval
$L_{\rm obs}^{\rm min} - L_{\rm model}^{\rm max}$.
We have verified
that all our results are not sensitive to the choice of the bin size.

The values of $\Xi^{\rm model}(z)$, evaluated
at all redshifts, 
are shown in Table \ref{tab1} for all models explored.
To compare these values with theoretical expectations, we use Monte
Carlo
techniques to
compute distribution of $\Xi$, $f(\Xi,z)$,  expected when 
$n_{\rm model}({\Delta L_i},z)$ is a Gaussian random variable,
normally distributed around
$n_{\rm obs}({\Delta L_i},z)$ with variance $10^{(\sigma_{\rm model}^2+
\sigma_{\rm obs}^2)}$.
This function is used to evaluate the cumulative 
probability of  $\Xi$ by integrating $f
(\Xi,z)$:
\begin{equation}
 P(\Xi^{\rm model},z)=\int_{>\Xi^{\rm model}(z)}{\rm d}\,\Xi f(\Xi,z)
 \label{eqn:P_M}
\end{equation}
which is defined in analogy to the $\chi^2$-probability and
represents the probability that a function that genuinely describes the
dataset would give a value larger or equal to $\Xi^{\rm model}$.
The values of $P(\Xi^{\rm model},z)$ are listed (in parentheses) in
Table \ref{tab1}
and indicated in each plot.

The results confirm those of the M06 analysis, 
in the sense that all models, apart from E1 that 
significantly overpredicts the abundance of AGNs at $z=2$, 
match the LFs in the range $1<z\le 2$ fairly well. 
The advantage of considering bolometric 
rather than B-band or hard X-ray LFs
is apparent at lower redshifts where discrepancies between models and 
observations at the bright and faint ends of the luminosity functions 
are more significant here than in the M06 analysis.
Indeed, all models overpredict the abundance of bright objects
and underpredict the abundance of the faint ones at $z=0.5$ and $z=1$.

In the LF bright end, the mismatch can be reduced by advocating some  
physical mechanism, like inefficient cooling, that hampers mass accretion in 
large haloes. Our simple model E2, in which mass accretion is inhibited in haloes 
with masses larger than $10^{13.5} M_{\odot}$, provides a better match to 
data, especially at $z=0.5$, although the effect is less apparent here than in 
M06 which considered the optical B-band LF. 
The overabundance of bright AGNs is also alleviated in model H
since the variable Eddington ratio guarantees that a BH hosted in the 
largest halos accretes most of the time at a sub-Eddington rate, resulting 
in a fainter AGN.

In all models, but E1, the LF faint end is biased low. The effect is systematic 
and, in the luminosity range accessible to observations, it does not depend on luminosity.
Discrepancies grow larger when extrapolating the comparison below
to objects fainter than  $L_{\rm bol}^{\rm min}$, below which the LF predicted 
by most semi-analytic models turns-over while the model LF of H06 is fitted by
a power-law. The power-law behaviour is, however, recovered by model H, that
assumes a time-dependent Eddington ratio.

\subsection{The hard X-ray luminosity function at $z \sim 0.1$}

\begin{figure}
\includegraphics[width=0.45\textwidth]{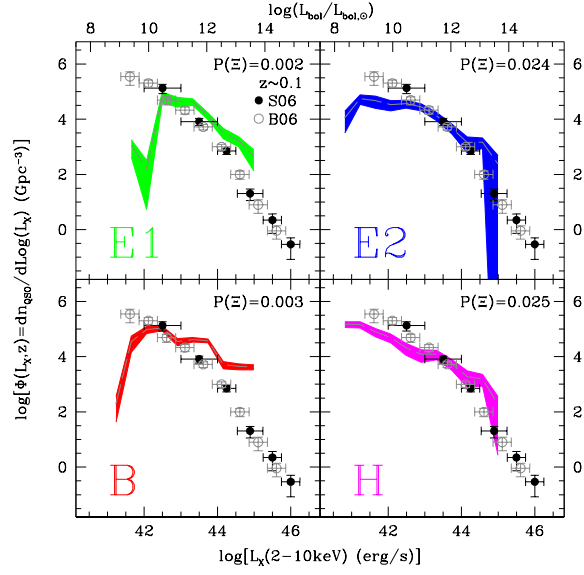}
\caption{
The AGN bolometric LF at $z=0.1$: models vs. observations.
S06 and B06 LFs are represented by filled and open dots, respectively.
Vertical error bars represent $1\sigma$ uncertainties while horizontal bars 
indicate the size of the luminosity bins.
Each plot refers to a different model, as indicated by the labels.
}
\label{fig:lf2}
\end{figure}

To understand whether the under-abundance of faint AGNs predicted by most semi-analytic models
is real or a mere artifact resulting from having extrapolated the power-law behaviour 
of the bolometric LF of H06 below $L_{\rm bol}^{\rm min}$
requires probing the AGN LF to lower luminosities, which is only possible in the nearby universe. 

In this section we do not compare the model LFs with the bolometric LF at $z\sim0$. Instead, we 
apply the inverse bolometric conversion of H06 to compare model predictions with the 
LFs of S06 and B06 at $z\sim 0.1$ in the $[2-10\,{\rm keV}]$ band.
The rationale behind this choice is as follow. 
First of all, these two datasets, especially the B06 one, include objects that were not considered in the H06 
analysis. 
Secondly, selection in the hard X-ray allows to include obscured AGNs which
make bolometric corrections less severe in this band.
Third, the two samples have rather similar composition as $90 \%$ of
the objects are Seyfert galaxies.
As a result, comparing 
model with S06 and B06 LFs allows to 
maximize the number of nearby, homogeneous objects, while 
reducing uncertainties in model bolometric corrections.

Model vs. data comparisons are shown in Fig. \ref{fig:lf2}, where the S06 and B06 LFs 
are represented by filled and open dots, respectively.
Vertical errorbars represent $1\sigma$ uncertainties, while horizontal bars 
indicate the size of the luminosity bins. 
The AGN luminosity in the B06 sample are measured in the $[20-40\,{\rm keV}]$ band and 
transformed in the $[2-10\,{\rm keV}]$ band according to 
$L_{[2-10\rm\,{\rm keV}]}/L_{[20-40\rm\,{\rm keV}]}=2.3$ \citep{beckmann2006}.

The shaded areas show the model LFs at $z=0.1$ together with their $1\sigma$ uncertainties
expressed in the $[2-10 {\rm keV}]$ band by using the bolometric correction of H06
\begin{equation}
  \frac{L}{L_{[2-10\rm\,{\rm keV}]}}=c_1\left(\frac{L}{10^{10}L_{\sun}}\right)^{k_1}+
  c_2\left(\frac{L}{10^{10}L_{\sun}}\right)^{k_2} \, ,
  \label{eqn:bolcorr}
\end{equation}
with s$c_1=10.83$, $k_1=0.28$, $c_2=6.08$ and $k_2=-0.02$.
The bolometric luminosities are indicated on the X-axis 
in the upper part of the plot. 
In order to correct for the extinction in the X-ray bands, we have used the 
following equation, also provided by H06:
\begin{equation}
  \frac{\Phi(L_{[2-10\rm\,{\rm keV}]})}{\Phi(L_{\rm bol}[L_{[2-10\rm\,{\rm keV}]}])}=f_{46}
    \left(\frac{L_{\rm bol}[L_{[2-10\rm\,{\rm keV}]}]}{10^{46}\,{\rm erg\,s^{-1}}}\right)^{\beta} \, ,
    \label {eqn:obscuration}
\end{equation}
where $f_{46}=1.243$, $\beta=0.066$ and $L_{\rm bol}[L_{[2-10\rm\,{\rm keV}]}]$ is the
bolometric luminosity correspondent to $L_{[2-10\rm\,{\rm keV}]}$,
as given by the bolometric corrections of Eq. (\ref{eqn:bolcorr}).

The comparison between model and data confirm our previous extrapolation, since
the observed number density of faint AGNs with 
$L_X=10^{42}-10^{43} {\rm erg/s}$
is significantly larger than that predicted by all models,
as indicated by the sudden drop in the values of the  $P(\Xi^{\rm model})$
at $z=0.1$. This is due to the fact that, for a given value of $\Xi^{\rm model}$, the
$f(\Xi)$ distribution at redshifts $\geq0.5$  is more positively skewed than
at $z=0.1$, as we have verified.
Discrepancies are larger for models 
E1 and B, while models E2 and H provide a better match to  
the faint end of the local LF.
The sharp downturn in the E1 and B models is a robust feature
since the characteristic mass of halos populating the faint luminosity
bins ($\sim 10^{11.5}\,M_{\sun}$) is well above the mass resolution limit 
in our merger trees.

Note that the largest discrepancies are found in the faintest luminosity bin
which can only be probed by the B06 sample. 
With this respect, it is worth noticing that \citet{sazonov2004}
have used yet another dataset of hard X-ray selected AGNs to estimate 
the AGN LF down to $L_{[3-20\rm\,{\rm keV}]}\sim10^{41}{\rm erg/s}$.
Their LF is consistent with those of S06 and B06 down to the faintest 
objects. The sample of \citet{sazonov2004} consists of 95 AGNs 
in the $[3-20\rm\,{\rm keV}]$ interval at high galactic latitude 
serendipitously detected in the RXTE slew survey. However, only 
$60 \%$ of these AGNs are classified as Seyfert galaxies, many of
which also belong to the S06 sample. Since in this work we prefer to deal with 
a homogeneous sample of local AGNs, we have decided not to consider 
the \citet{sazonov2004} LF in our quantitative analysis.

\subsection{The biasing function}

\begin{figure}
\includegraphics[width=0.45\textwidth]{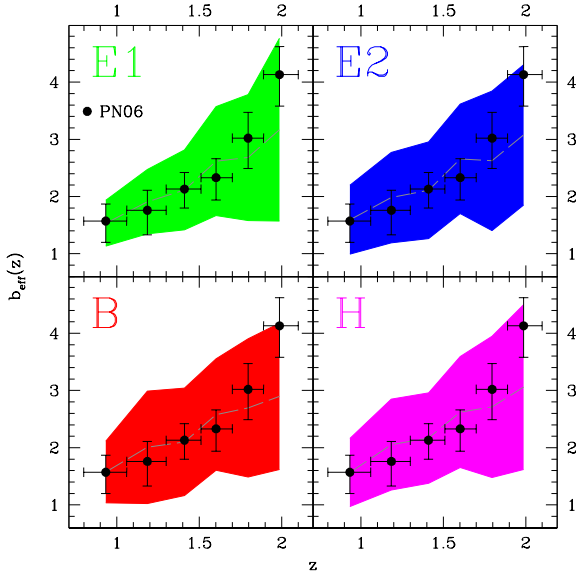}
\caption{
The AGN bias function at $z<2$: models vs. observations. 
The solid black points show the bias of \citet{porciani_norberg2006}.
The shaded areas show the bias predicted by our four models.
The four dashed lines, with their shaded areas,
show our model predictions with their $1\sigma$ uncertainties.
}
\label{fig:bias}
\end{figure}

For the sake of completeness, we follow the M06 analysis and
quantify the clustering of our model AGNs at $z\le2$ 
through their biasing function, $b_{\rm eff}(z)$.
The latter has been computed by weighting the 
analytic biasing function of the DM haloes provided by \citet{sheth_mo_tormen2001},
$b(M_{\rm halo},z)$, with the mass function of the haloes hosting AGNs with
luminosities larger than the thresholds of the observations,$\Psi(M,z)$:
\begin{equation}
  b_{\rm eff}(z)=\frac{\displaystyle\int_{0}^{+\infty}b(M_{\rm halo},z)
    \Psi(M_{\rm halo}(L_B>L_{{\rm min},B}),z)dM_{\rm halo}}
  {\displaystyle\int_{0}^{+\infty}\Psi(M_{\rm halo}
    (L_B>L_{{\rm min},B}),z)dM_{\rm halo}}\ ,
\end{equation}
where the minimum B luminosity at the 5 redshifts explored is
$L_{{\rm min},B}/L_{\odot,B}(z) =\{1.42\cdot10^{11},3.58\cdot10^{11},
3.92\cdot10^{11},6.82\cdot10^{11},6.82\cdot10^{11},1.08\cdot10^{12}\}$.
The difference with respect to the M06 analysis is that here we 
have used the bolometric correction of H06 to convert our bolometric luminosities to
B-band ones, i.e. using the Eq. (\ref{eqn:bolcorr}) with $c_1=6.25$, $k_1=-0.37$, 
$c_2=9$ and $k_2=-0.012$, and the Eq. (\ref{eqn:obscuration}) 
with $f_{46}=0.26$, $\beta=0.082$.
We have compared model predictions with the most recent
observational determination of the biasing function 
at $0\leq z\leq2$ \citep{porciani_norberg2006}, estimated in the B-band.

The result is shown in Fig. \ref{fig:bias}, where the points represent the
observed B-band AGN biasing  function and the shaded areas 
show the $1\sigma$ uncertainty interval around  model predictions.
As in the M06 analysis, the large model uncertainties
do not allow us to place strong constraint based on the AGN clustering.
Indeed, all our models are in acceptable agreement with the data,
suggesting, however, that possible disagreements may become significant  
at $z > 2$.

\section{Discussion and conclusions}
\label{sec:disc_concl}
 
In this paper we tested the validity of the assumption 
that the evolution of AGNs is simply related to the cosmological merging history of DM haloes.
To do that, we have compared the predictions of hierarchical semi-analytic models 
with the most recent determination of the AGN LF in the hard X-ray band and 
their biasing function in the B-band at $z \le 2$.
Our main results can be summarized as follows.

(i) We confirm the success of simple semi-analytic models in reproducing both 
the AGN bolometric LF at $1<z\leq2$, i.e. around the peak of activity, and
their clustering, quantified by the biasing function, at $z\leq 2$.

(ii) As pointed out by several previous analyses, problems occur at lower redshifts, where hierarchical models 
systematically overestimate the number density of bright AGNs and underestimate the faint ones.

(iii) Comparing bolometric LFs rather than the optical or hard X-rays ones
allows to spot significant discrepancies already at moderate redshifts $z\sim1$, i.e. 
earlier than what was found in previous analyses (e.g. M06).
     
(iv) The predicted number density of bright AGNs can be reduced not only by advocating 
inefficient cooling within massive haloes, as in model E2, but also by accounting for 
feedback mechanisms, as we did in model H.

(v) The underestimate of faint AGNs looks like a more serious problem that we have tried to tackle by
assuming a time-dependent Eddington ratio, as suggested by the outcome of
the hydrodynamical simulations by \citet{hopkins2005}. As shown by
\citet{volonteri_salv_haardt2006}, implementing this prescription within 
a semi-analytic framework, as we did in model H, proved to 
be successful in reproducing the redshift distribution of the faint X-ray counts \citep{volonteri_salv_haardt2006}.
In this work, we extended the analysis of \citet{volonteri_salv_haardt2006}
by comparing  model predictions with the most recent determinations of the local AGN LF by S06 and B06
in the hard X-ray band, to include absorbed AGNs and minimize the impact of bolometric corrections.
This is a very demanding test for semi-analytic models,  
which constitutes the main focus of this paper since, as we have pointed out,
the mismatch in the number density of faint AGNs grows larger when decreasing the redshift.
We found that the two most successful models E2 and H are in acceptable agreement with the data at $z\gtrsim 0.5$, 
but struggle to match the correct number density of faint X-ray sources in the nearby universe.

\begin{figure}
\includegraphics[width=0.45\textwidth]{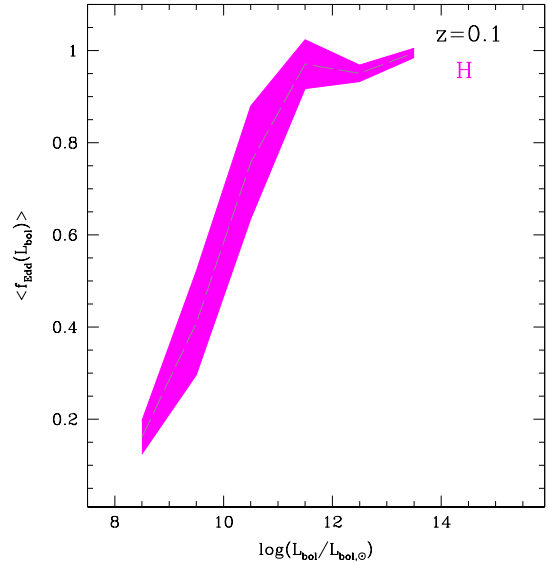}
\caption{
The mean Eddington ratio in function of the AGN luminosity,
at $z=0.1$, for the model H. The coloured area represents the
$1\sigma$ uncertenties.
}
\label{fig:f_Edd}
\end{figure}

Model H, based on the results of hydrodynamical simulations of \citet{hopkins2005} within a pure merger driven scenario, 
seems unable to account for local faint AGNs. If the accretion efficiency were much lower, 
the lifetime of faint AGNs would increase proportionally and help alleviate the discrepancy. 
However, the Eddington factors derived from \citet{hopkins2005} light curve are well below 
$f_{\rm Edd}=0.1$ only when a galaxy hosts a black hole with an initial mass anomalously smaller than that  
predicted by the  $M_{\rm BH}-\sigma_g$ correlation. This is evident in Figure \ref{fig:f_Edd}: our models  
assume that accretion processes bring the black holes onto the  $M_{\rm BH}-\sigma_g$ relation 
and the accretion efficiency is for most systems above $f_{\rm Edd}=0.1$ . This can be understood 
using a very simple argument. Let us assume that (i) quiescent black holes sit on the $M_{\rm BH}-\sigma_g$ relation, 
as observed in the nearby galaxy where the $M_{\rm BH}-\sigma_g$ relation was indeed derived. 
This is therefore a safe assumption in the local Universe.  
(ii) Accretion is triggered only by major mergers, that is mergers between similar size galaxies, 
with a mass-ratio larger than at least 0.1 \citep{cox2004}.  
And, (iii) an accretion episode grows black holes until they reach the $M_{\rm BH}-\sigma_g$ 
relation for the newly formed galaxy, due to feedback effect.  Studies of local samples of AGN 
seem indeed to confirm that typically AGN masses scale with the  $M_{\rm BH}-\sigma_g$ relation \citep[e.g., ][]{ferrarese2001, greene2006}. 
Within these simple but sensible assumptions, the accretion efficiency is bound to be high, 
as can be easily estimated by equation \ref{eq:fedd}. If we consider, for example, a major merger 
of a Milky-Way sized galaxy, the Eddington factor of the black holes remains $f_{\rm Edd}<0.1$ for only about $10^6$ yr. 

The inadequacy of the pure merger driven scenario becomes more evident when considering the 
observational constraints on the Eddington factor of Seyfert galaxies, which constitute about $90 \%$ of the local AGN population. 
 \citet{woo2002} analyze a sample of 234 AGNs at $0.001<z<1$, composed, at $z\leq0.1$ mainly by Seyfert galaxies. 
They find a large scatter (2 orders of magnitude) in the Eddington factor at both fixed luminosity and fixed BH mass. 
\citet{woo2002} do not find any trends of the Eddington factor with either luminosity, mass or redshift, 
which cannot be explained by selection effects.

It turns out that the S06 and B06 catalogues are largely composed by Seyfert galaxies that
constitute respectively $94 \%$ and $88 \%$ of the total galaxy host populations.
Only a small fraction of these local Seyfert galaxies have disturbed morphology,
and thus did not experience any recent merging event. Indeed,
only $4 \%$ of the sources in the S06 catalog are hosted in galaxies 
that show evidences of recent interactions. The AGNs in the B06 catalogue are 
typically found at low galactic latitudes which hamper a systematic analysis of their
host galaxy morphology. Yet, the similar galaxy composition of the
two catalogues suggests that also B06 AGNs preferentially populate quiescent environments.
Based on this observational evidence, it may be suggested that galaxy mergers might not constitute the 
only trigger to AGN activity.

To decide whether this is indeed a viable hypothesis, it is worth reviewing the 
observational evidences of local AGN samples.
Bright,  low-redshift quasars and ultra luminous infrared
galaxies, that are generally regarded as hosting obscured AGNs,
are often found in merging systems 
\citep[see, e.g.][and references therein]{sanders1996,canalizo2001,capetti2006}
which indicates a possible connection between mergers of gas-rich galaxy and AGN activity.
On the contrary, as we have seen, fainter AGNs typically reside in quiescent, non-interacting galaxies
\citep[e.g.][and references therein]{kauffmann2003,grogin2005}.   
However, this alone does not guarantee that an alternative AGN triggering mechanism
is at work, as this observational evidence can still fit into a merger-driven scenario.
In fact, the brightest among these objects could be the relics of a previous bright quasar 
epoch in a spheroid-forming merger \citep[see, e.g.][and references therein]{hopkins2006_2},
while the fainter ones would consist of AGNs hosted in
``dead'' elliptical galaxies fueled via accretion of hot 
spheroid gas and steady mass loss from stars \citep[see, e.g.][]{ciotti2001,sazonov2005,croton2006},
an accretion mode which cannot dominate the BH growth.

The merger-driven scenario, however, proved to be inadequate in accounting for 
the relatively high accretion rate AGNs
observed at low redshifts in undisturbed, late-type, star-forming
galaxies with low mass ($\lesssim 10^{7}\,M_{\sun}$) BHs \citep[e.g.][]{kauffmann2003}. 
Indeed alternative mechanisms, not included in our simple models, 
have been suggested to trigger the mass accretion in these objects.
For instance, it has been proposed that a significant contribution to the faint AGN mass accretion 
could come from the material liberated by the tidal disruption of stars
by the central BHs \citep{milosavljevic2006}, or by the mass 
of the stars captured by the BH disks and eventually dissolved \citep{miralda2005}.  
Other studies have considered the stochastic accretion of molecular clouds in quiescent systems
\citep[see e.g.][]{hopkins_hernquist2006,croton2006}. 
Moreover, it was suggested that also disk instability could trigger mass accretion,
contributing to increase even more the number density of faint AGNs 
\citep[see e.g.][]{croton2006,bower2006}. Finally, a better treatment of mergers between haloes
with low mass ratio may also contribute to solve these problems \citep[see e.g.][]{malbon2006,croton2006}.

Whether including these alternative trigger mechanisms in our simple merger-driven scenario
can help in reconciling model predictions with observations at $z\sim0$ is a question that
deserves further investigation.

\section*{Acknowledgements}
FM thanks the Institute of Astronomy, University of Cambridge, for the kind hospitality.
This work has been partially supported by ASI and INAF.
We would also like to thank the anonymous referee for very useful comments.

\bibliographystyle{mn2e}
\bibliography{master_qso}

\label{lastpage}
\end{document}